\documentclass[dvipdfmx,aps,prb,twocolumn,superscriptaddress,longbibliography,10pt]{revtex4-1}

\usepackage{amsmath} 
\usepackage[dvipdfmx]{graphicx}
\usepackage{verbatim} 
\usepackage{color} 
\usepackage{subfigure} 
\usepackage{here}

\usepackage{algorithm}
\usepackage{algpseudocode}

\usepackage{physics}

\usepackage{pgfplots}
\usepackage{tikz}

\tikzset
 {every pin/.style = {pin edge = {<-}}, 
 > = stealth,       
 flow/.style = 
 {decoration = {markings, mark=at position #1 with {\arrow{>}}},
 postaction = {decorate}
 },
 flow/.default = 0.5,   
 main/.style = {line width=1pt}     
 }

\newcommand{\rr}{\boldsymbol{r}}
\newcommand{\kk}{\boldsymbol{k}}

\newcommand{\zzt}{\boldsymbol{\zeta}}

\newcommand{\de}{\mathrm{d}}
\newcommand{\e}{\mathrm{e}}
\newcommand{\B}{\mathrm{B}}

\newcommand{\beq}{\begin{eqnarray}}
\newcommand{\eeq}{\end{eqnarray}}
\newcommand{\del}{\partial}

 \renewcommand{\algorithmicrequire}{\textbf{Input:}}
 \renewcommand{\algorithmicensure}{\textbf{Output:}}

 \makeatletter
\let\Hy@backout\@gobble
\makeatother
 
\begin{document}

\title{A Phase Prediction Method for Pattern Formation in\\ Time-Dependent Ginzburg-Landau Dynamics for Kinetic Ising Model\\ without \textit{a priori} Assumptions on Domain Patterns}
\author{Ryoji Anzaki}
\affiliation{Earthquake Research Institute, The University of Tokyo, 1-1-1, Yayoi, Bunkyo-ku, Tokyo 113-0032, Japan}
\author{Shin-ichi Ito}
\author{Hiromichi Nagao}
\affiliation{Earthquake Research Institute, The University of Tokyo, 1-1-1, Yayoi, Bunkyo-ku, Tokyo 113-0032, Japan}
\affiliation{Graduate School of Information Science and Technology, The University of Tokyo, 7-3-1, Hongo, Bunkyo-ku, 113-8656 Tokyo, Japan}
\author{Masaichiro Mizumaki}
\affiliation{Japan Synchrotron Radiation Research Institute (JASRI/SPring-8), 1-1-1 Kouto, Sayo, Hyogo 679-5198, Japan}
\author{Masato Okada}
\affiliation{Graduate School of Frontier Sciences, The University of Tokyo, Kashiwa, Chiba 277-8561, Japan}
\author{Ichiro Akai}
\affiliation{Institute of Pulsed Power Science, Kumamoto University, 39-1 Kurokami 2, Kumamoto 860-8585, Japan}

\date{\today}

\begin{abstract}
We propose a phase prediction method for the pattern formation in the uniaxial two-dimensional kinetic Ising model with the dipole-dipole interactions under the time-dependent Ginzburg-Landau dynamics. Taking the effects of the material thickness into account by assuming the uniformness along the magnetization axis, the model corresponds to thin magnetic materials with long-range repulsive interactions. We propose a new theoretical basis to understand the effects of the material parameters on the formation of the magnetic domain patterns in terms of the \textit{equation of balance} governing the balance between the linear- and nonlinear forces in the equilibrium state. Based on this theoretical basis, we propose a new method to predict the phase in the equilibrium state reached after the time-evolution under the dynamics with a given set of parameters, by approximating the third-order term using the restricted phase-space approximation [R. Anzaki, K. Fukushima, Y. Hidaka, and T. Oka, \textit{Ann. Phys.} \textbf{353}, 107 (2015)] for the $\phi^4$-models. Although the proposed method does not have the perfect concordance with the actual numerical results, it has no arbitrary parameters and functions to \textit{tune} the prediction. In other words, it is a method with no \textit{a priori} assumptions on domain patterns.
\end{abstract}

\maketitle

\section{Introduction}
Magnetic materials are of great interest even before the beginning of the application of quantum physics to the solid-state physics \cite{vleck1945survey}. The domain patterns are essential for understanding the magnetic materials since the macroscopic properties of magnetic materials are largely affected by the domain patterns \cite{kittel1949physical}. In the light of recent progress in experimental methods to observe the magnetic domain patterns, it is now convincing that one may obtain information on the magnetic dynamics, e.g., the material parameters, the external magnetic fields, and the size of the magnetic materials from the domain patterns. Among such experimental methods, X-ray magnetic circular dichroism (XMCD) \cite{suzuki2013hard} and the detection of the Kerr effect \cite{argyres1955theory} via visible light \cite{reif1991effects} are well-known methods to detect the magnetization normal to the surface of the materials. In the realm of theoretical- and simulation physics, researchers have already made progress towards this aim. Jagra \cite{jagra2004numerical} and Kudo \textit{et al}. \cite{kudo2007field} performed numerical simulations using similar models to reproduce the magnetic domain patterns on two-dimensional magnetic materials. The latter proposed a relation between the sweep rate of the external magnetic field and the final magnetic domain patterns in the equilibrium state. They utilized the two-dimensional kinetic Ising spin system with spins on the square lattice lying on the $xy$-plane, while the magnetization is restricted in the $z$-direction, which is normal to the $xy$-plane. Assuming that the high-wavenumber components of the Green's function of the dipole-dipole interaction play few roles, they succeeded in explaining the various domain patterns resulting from different sweep rates by solving the time-dependent Ginzburg-Landau (TDGL) equation numerically \cite{kudo2007field}. Iwano \textit{et al}. \cite{iwano2014maze} adopted a numerically evaluated effective two-dimensional Green's function for the dipole-dipole interaction.

In the early history of the researches of the TDGL dynamics, the probability density functions (PDFs) of the spin systems under the TDGL dynamics have been studied by Kawasaki \cite{kawasaki1974macroscopic, kawasaki1974contributions, kawasaki1974contribution} in the 1970s. Suzuki \textit{et al}. \cite{suzuki1973calculation} also studied the same system using the Markov chain. Their major interests were to obtain the global characteristics of the spin configuration, e.g., the dynamic magnetic susceptibility \cite{kawasaki1974macroscopic} and the critical exponents \cite{suzuki1973calculation}, using the analytical tools including the diagrammatic methods. In the 1980s, Grant \textit{et al}. \cite{grant1985theory} investigated the similar system in a context of the phase separation, and developed a theory using the spatial wavenumbers of the fields. On the other hand, Kawasaki \cite{kawasaki1982kink} also investigated the kink dynamics in the one-dimensional TDGL model, whose achievements have been inherited to the researches on the dynamical phase-transition in the TDGL dynamics of the XY-model \cite{yasui2002dynamic, fujiwara2004magnetic}. In the late 1980s, numerical simulations have been performed using the TDGL equation in the real-space \cite{rogers1988numerical}. 

In the realm of magnetic materials, the explanations of the magnetic domain patterns have been developed for decades \cite{kittel1946theory, kaplan1993domain, bochi1995magnetic}. 
The Kooy-Enz model \cite{kooy1960} and its variants \cite{kaplan1993domain, lisfi2002magnetic} assume simple domain patterns specified by functions with one or more parameters and minimize the total energy (the sum of the contributions from the domain and the domain wall) with respect to the parameters. The forms of the functions that determine the domain patterns are chosen \textit{a priori} so that the entire problem simplifies into an optimization problem of real functions. Garel \textit{et al.} \cite{garel1982phase} analyzed the behavior of the similar system under finite temperature $T$ and external magnetic field $H$ thermodynamically and plotted the $T$-$H$ phase diagram with three phases named \textit{uniform, bubble}, and \textit{striped}. These phases are also defined by simple analytic functions with a few parameters.

In this Paper, we take a new strategy that does not involve any \textit{a priori} defined functions. The effects of the material thickness and other parameters to the TDGL pattern formation are explained by the newly proposed \textit{equation of balance} that describes the balances between the linear- and nonlinear forces in the equilibrium state reached after appropriate numerical time-evolutions with a realistic initial condition. This equation enables us to predict the phase that a specific TDGL equation with a given set of parameters forms in the equilibrium state. In the language of the magnetism, we can predict the pattern of the magnetic domain formed in thin magnetic materials for a given set of the TDGL parameters with the proposed method.

In this Paper, we use a numerical method to construct an effective two-dimensional Green's function by analytically averaging the dipole-dipole interactions along the $z$-direction for each grid point on the $xy$-plane as proposed in Ref. \cite{iwano2014maze}, enabling us to take the effects of the thickness into account more precisely.

\section{Model and Methods}
We utilize the Ising-like spin model with TDGL dynamics \cite{grant1985theory, rogers1988numerical, kudo2007field, kudo2007magnetic, yokota2017three}, also referred as the \textit{kinetic Ising model} \cite{tome1990dynamic}. We prepare an array of complex variables $\{\phi(\rr)\}$ with $\rr$ being an element of two-dimensional discrete space $\mathcal{D} = \{(x,y) : 1\leq x,y \leq L \} \subset \mathbf{Z}^2$ for a positive integer $L$. Each variable $\phi$ is regarded as a magnetic dipole restricted in the $z$-direction, while the vector $\rr$ represents a coordinate on the $xy$-plane, normal to the $z$ axis. Note that $x$- and $y$-components of spins are set to zero in this model. Introducing the saturation magnetization $\rho > 0$, the TDGL equation for the spin system above with time parameter $t$ is,
\beq\label{eq:Model:EOM}
\dv{\phi(\rr)}{t} = W(\rr| \phi] + B(t),
\eeq
where $B$ is the explicitly time-dependent external magnetic field (restricted in the $z$-direction), and $W(\rr| \phi]$ is a function of $\rr$ and a functional of $\phi$, defined as
\beq\label{eq:Model:Force}
W(\rr| \phi] = 
\alpha[\phi(\rr) - \rho^{-2}\phi^{3}(\rr)] + \beta\nabla^{2}\phi(\rr) - \gamma F[\phi],\\
F[\phi] = \int\de^{2} r' G(\rr-\rr')\phi(\rr').
\eeq
The terms containing $\alpha$, $\beta$, and $\gamma$ correspond to the anisotropy-, exchange- and the dipole-dipole interactions, respectively. The last term is represented via the Green's function for the magnetic dipole-dipole interaction $G(-)$.

By moving into the wavenumber space by the (non-unitary) Fourier transform
\beq\label{eq:model:fft:def}
\ev{f}_{\kk} = L^{-2}\sum_{\rr\in\mathcal{D}}f(\rr)\e^{i\kk\cdot\rr},
\eeq
the TDGL above becomes,
\beq
\dv{\phi_{\kk}}{t} = W_{\kk}[\phi] + B_{\kk}(t),
\eeq
with $W_{\kk}[\phi], \phi_{\kk}$ and $B_{\kk}$ being the spatial Fourier transformation of $W(\rr|\phi], \phi(\rr)$ and $B(t)$. Performing the Fourier transform, one obtains
\beq\label{eq:Model:RHS}
W_{\kk}[\phi] = \alpha\ev{\phi - \frac{\phi^{3}}{\rho^2}}_{\kk} - \beta|\kk|^{2}\phi_{\kk} - \gamma L^{2}\cdot G_{\kk}\phi_{\kk}.
\eeq
Here, the Fourier transformation of $G$ is introduced via the convolution theorem, and the prefactor $L^{2}$ is due to the choice of the Fourier transform Eq.~(\ref{eq:model:fft:def}). 

The effects of thickness are not apparent but introduced via the Fourier transform of the Green's function of the dipole-dipole interaction $G_{-}$ as already performed in \cite{iwano2014maze}. Hereafter, we assume that spin variables have the same value along with the $z$-direction for each $\rr$. The thickness (the spatial extension along the $z$-direction) of the material is assumed to take a positive value $A > 0$. By introducing the virtual $z$-coordinate $0\leq z\leq A$, we define an effective two-dimensional Green's function under the conditions specified above, as
\beq\label{Eq:Model:Green}
G(\rr;A) = \frac{1}{A^{2}}\int_{0}^{A}\de z\int_{0}^{A}\de z' \mathcal{G}(\rr;z-z'),
\eeq
with,
\beq
\mathcal{G}(\rr; \Delta z) = \frac{1}{(|\rr|^{2} + \Delta z^{2})^{3/2}} - \frac{3\Delta z^{2}}{(|\rr|^{2} + \Delta z^{2})^{5/2}}.
\eeq
The integral Eq.~(\ref{Eq:Model:Green}) can be performed analytically, and 
\beq
G(\rr;A) = \frac{2}{A^{2}}\left(\frac{1}{|\rr|} - \frac{1}{\sqrt{|\rr|^{2} + A^{2}}}\right).
\eeq
Note that in the limit $A\to0$, $G(\rr)$ converges to the inverse-cubic law point-wise. 

One may consider the \textit{continuum limit}, which corresponds to the case when the correlation length measured in the unit of the grid spacing becomes positive infinity. In that case, the Fourier transform of the Green's function is obtained from the real-space function $G(-)$ and has the analytical form 
\beq\label{eq:Model:Green:analytic}
G_{\kk}(A) = \frac{1}{\pi A^2}\frac{1-\e^{-A|\kk|}}{|\kk|}.
\eeq
The weight of the Fourier transformation is taken to be $(2\pi)^{-d}$, where $d=2$ is the spatial dimension. In this limit, the right-hand side of the equation of motion Eq.~(\ref{eq:Model:RHS}) becomes
\beq\label{eq:Model:continuum}\nonumber
W_{\kk}[\phi] = \alpha\ev{\phi - \frac{\phi^{3}}{\rho^2}}_{\kk} - \beta|\kk|^{2}\phi_{\kk} - \gamma (2\pi)^{2}G_{\kk}(A)\phi_{\kk}.\\
\eeq
Note that this representation is formally obtained simply by a replacement $L \to 2\pi$.

\section{Numerical Simulations}
In the $xy$-plane, we use the non-unitary fast Fourier transform (FFT) corresponding to Eq.~(\ref{eq:model:fft:def}) to construct the modes $\phi_{\kk} = \ev{\phi}_{\kk}$ and the wavenumber representation of the Green's function $G_{\kk}$. We adopt the periodic boundary condition for $x$- and $y$-direction, hence the entire topology of the simulation space is a torus. The spacing of the grid on the $xy$-plane is set to unity. We introduce a randomness of the coefficient of the anisotropy as $\alpha\to\alpha \Lambda(\rr)$ with $\Lambda(\rr) = 1 + \lambda(\rr)/4$ and Gaussian noise $\lambda(\rr) \sim \mathcal{N}(0, 0.3)$ independently and identically for all $\rr$, as in Ref. \cite{kudo2007field}. The external magnetic field intensity is represented by the rectified linear unit function $R(-)$ as $B(t) = R(B_0 - v_{\B}t)$ with $B_0, v_{\B} \geq 0$.

\begin{figure}
\begin{flushleft}
\hspace{0.05\linewidth} (a) Symmetric phase ($A = 1.0$)
\end{flushleft}
\vspace{-0.3cm}
\includegraphics[width=0.45\linewidth]{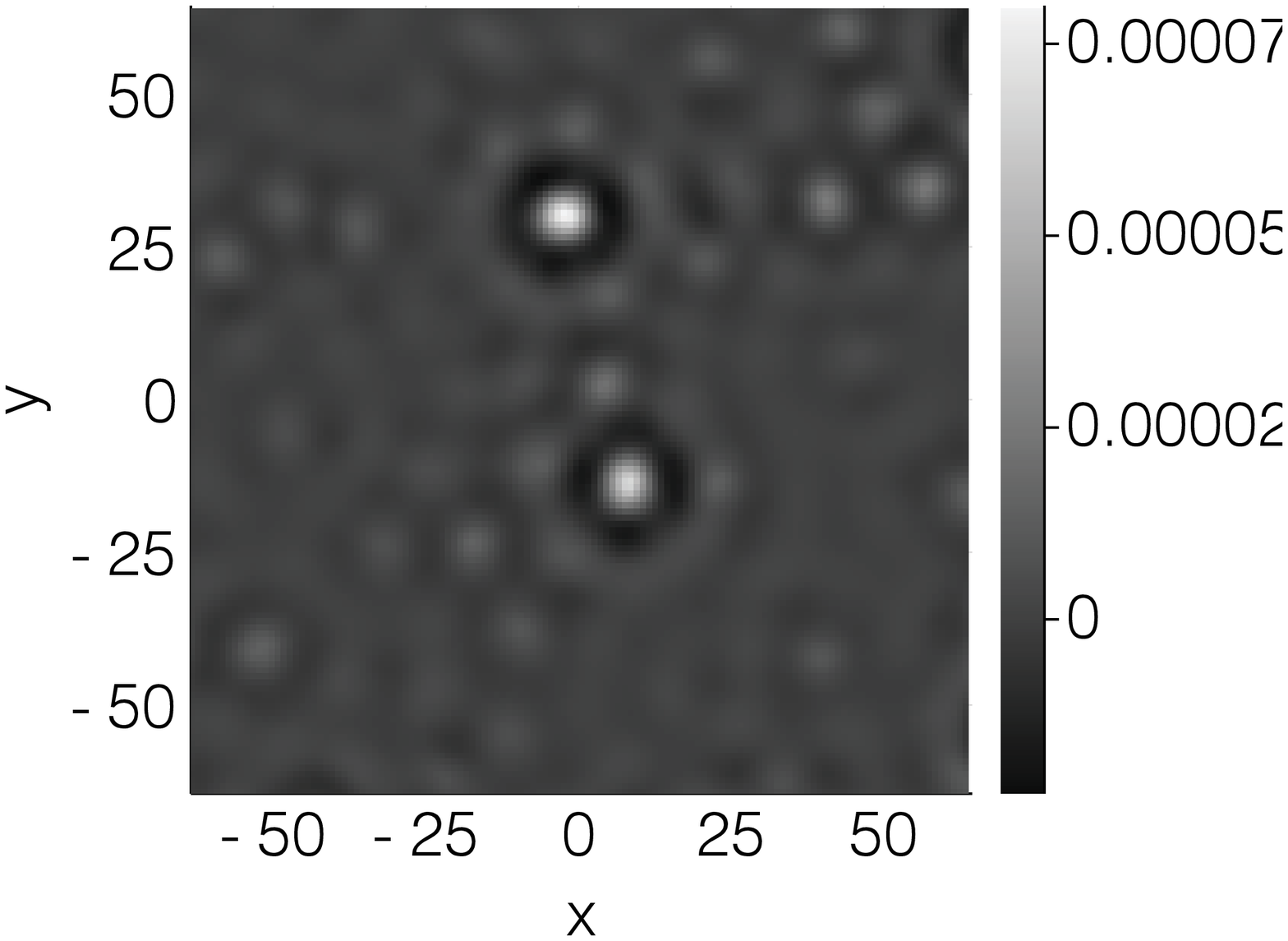}
\includegraphics[width=0.45\linewidth]{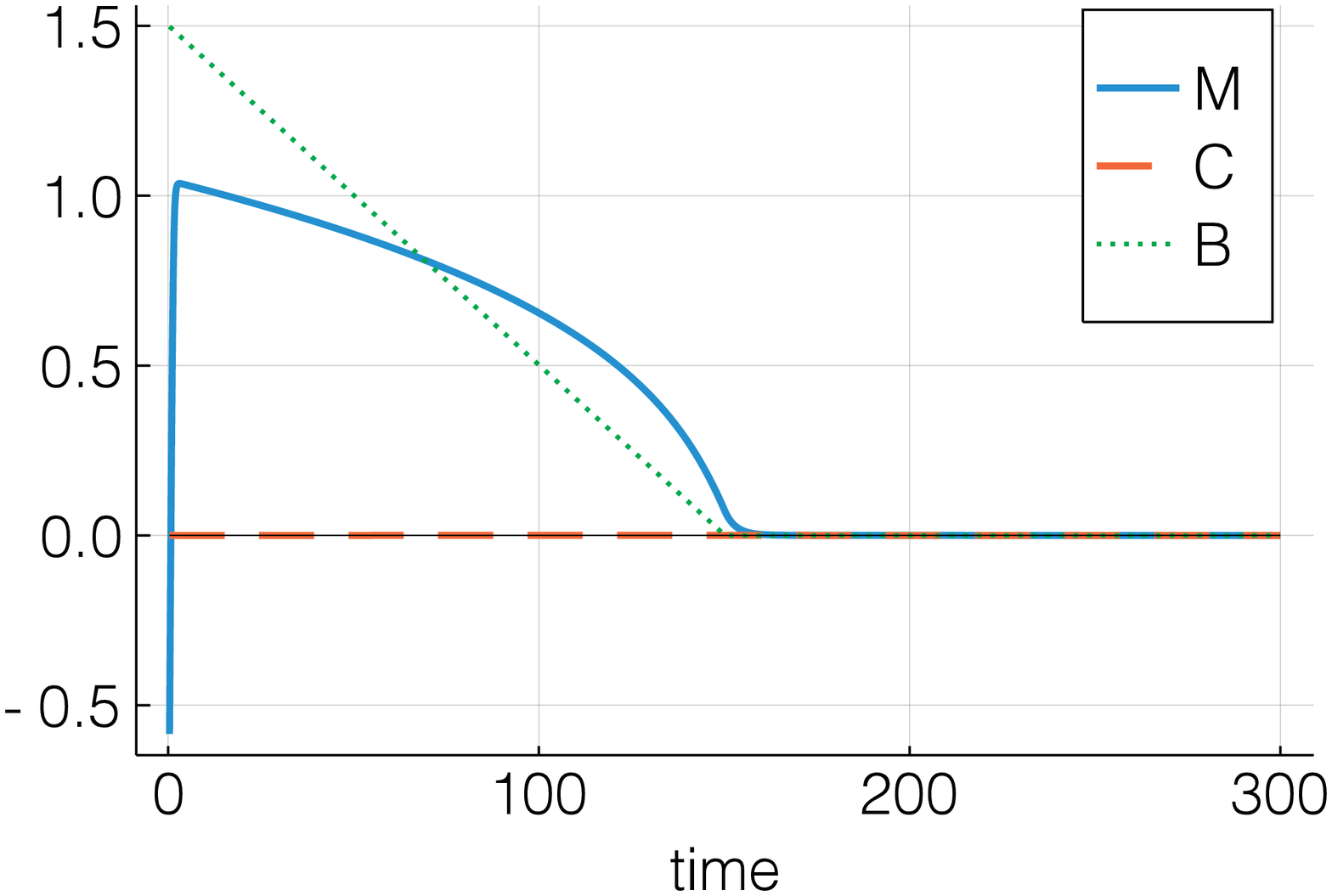}

\vspace{-0.4cm}
\begin{flushleft}
\hspace{0.05\linewidth} (b) T-breaking phase ($A = 1.5$)
\end{flushleft}
\vspace{-0.3cm}
\includegraphics[width=0.45\linewidth]{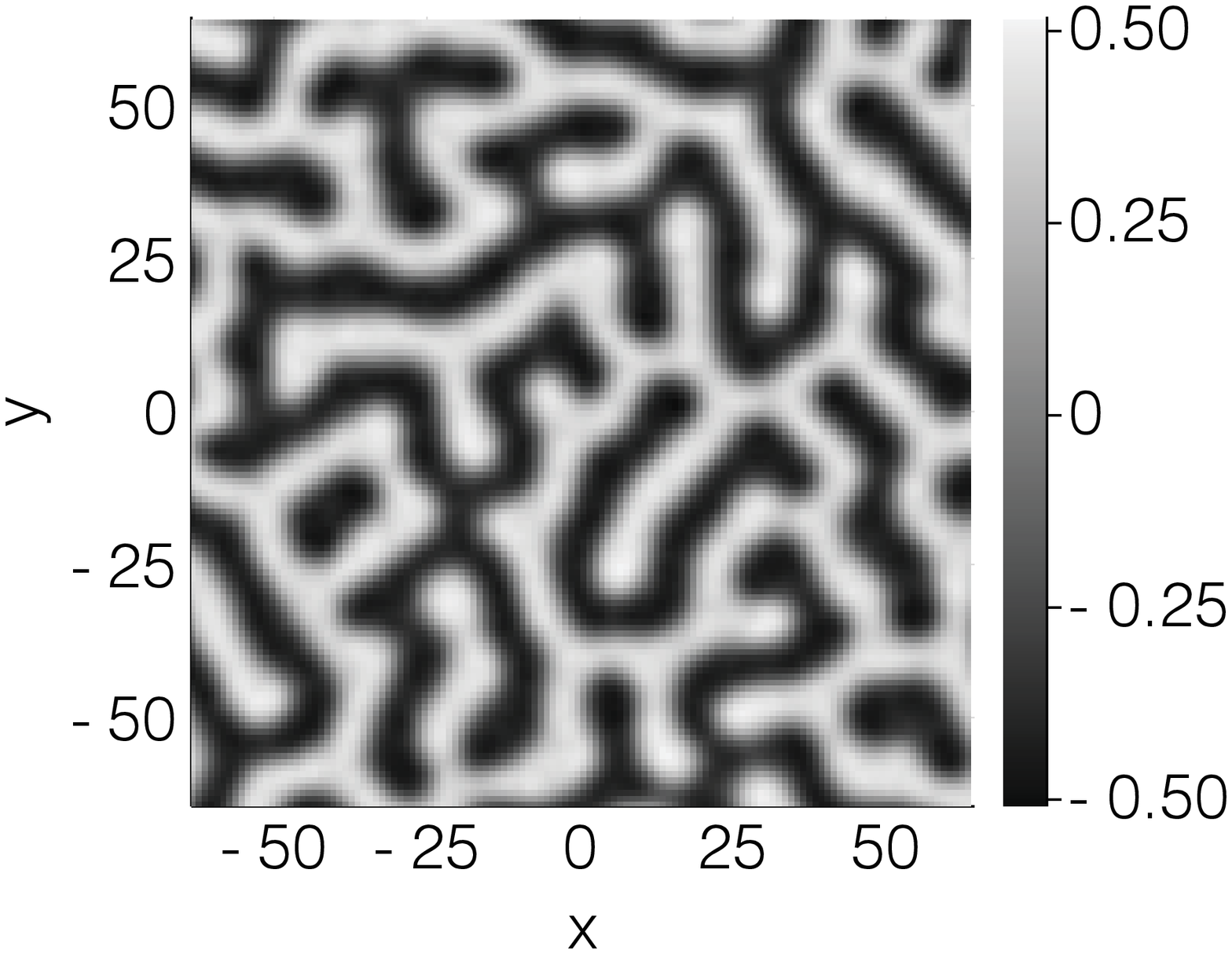}
\includegraphics[width=0.45\linewidth]{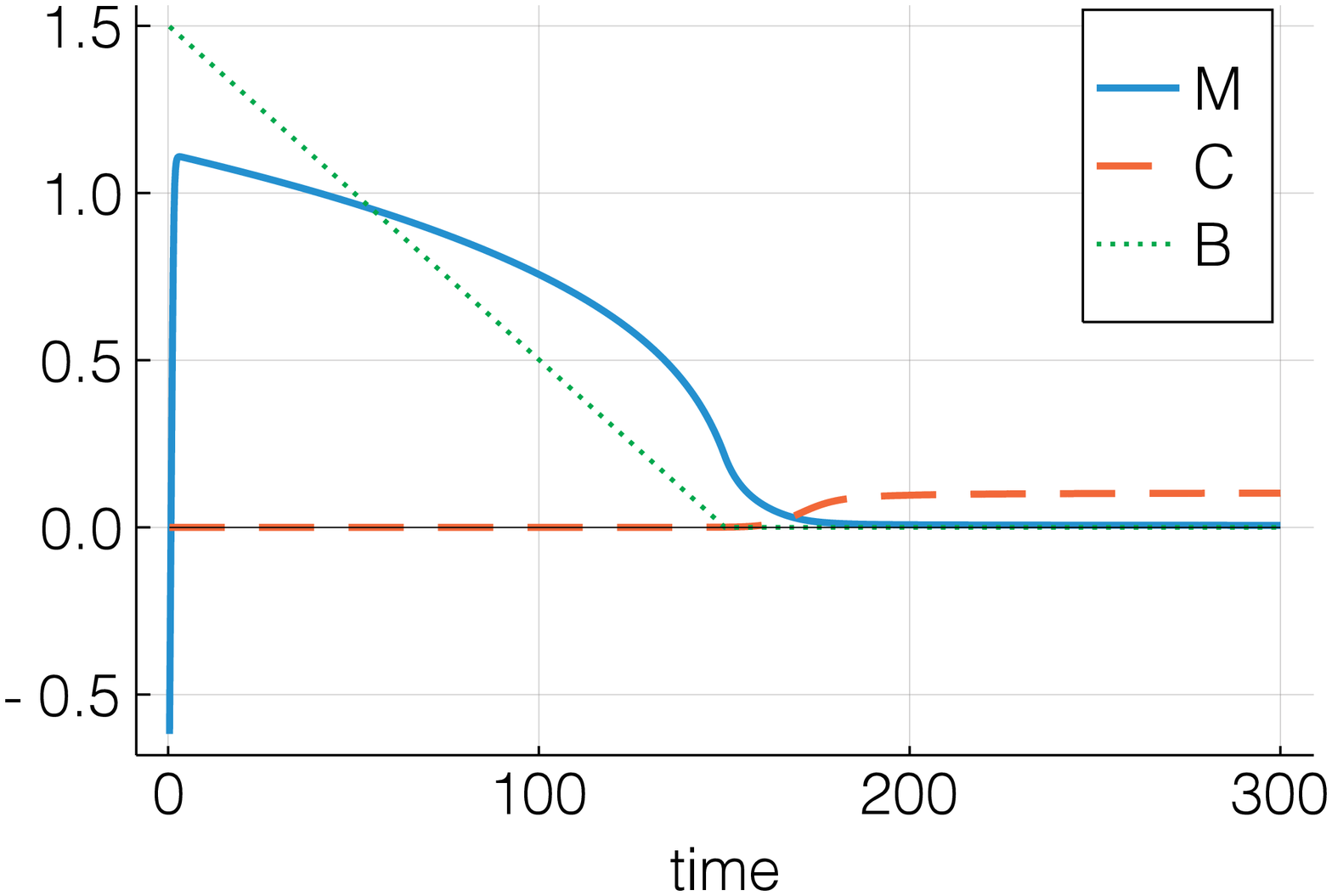}

\vspace{-0.4cm}
\begin{flushleft}
\hspace{0.05\linewidth} (c) TZ-breaking phase ($A = 2.0$)
\end{flushleft}
\vspace{-0.3cm}
\includegraphics[width=0.45\linewidth]{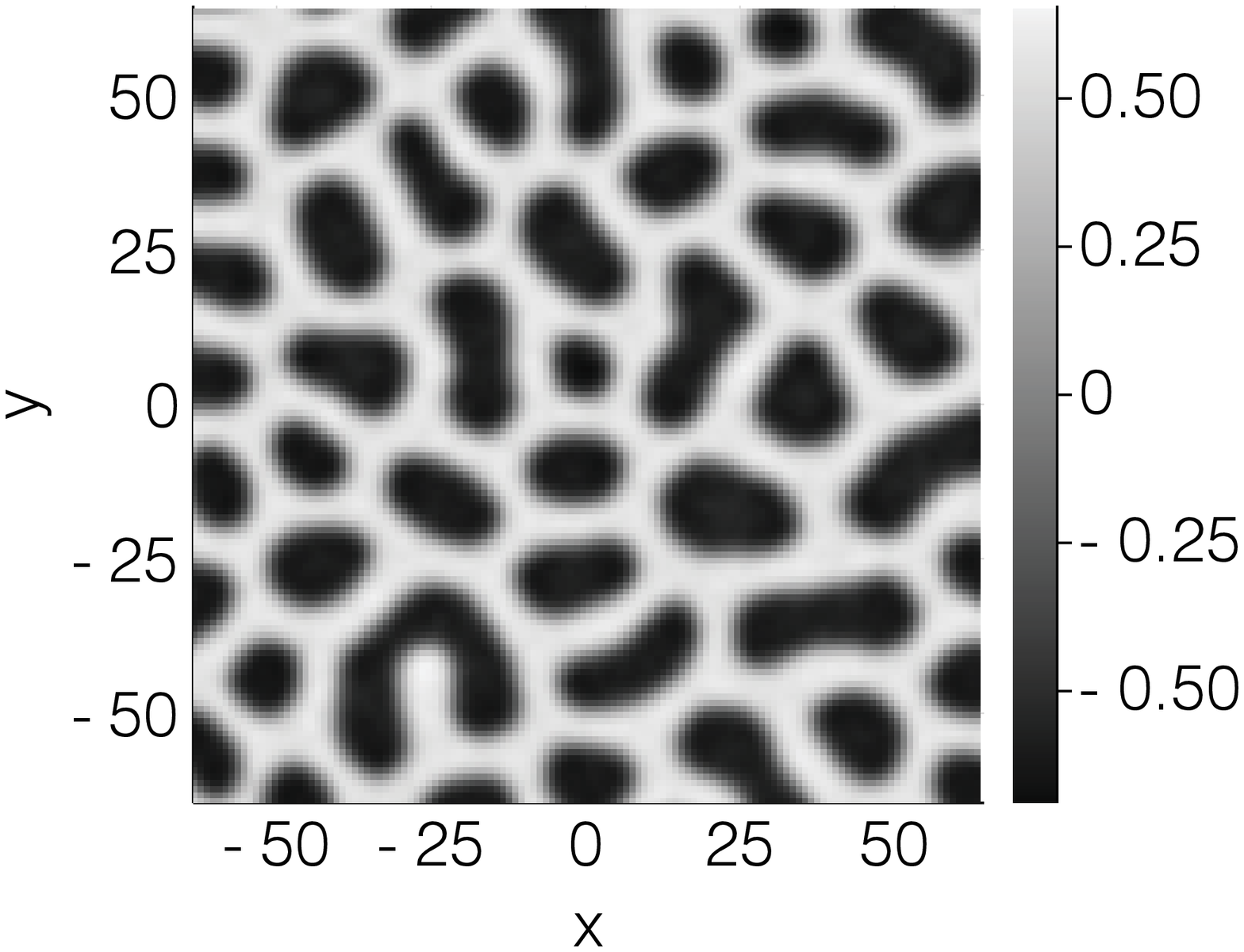}
\includegraphics[width=0.45\linewidth]{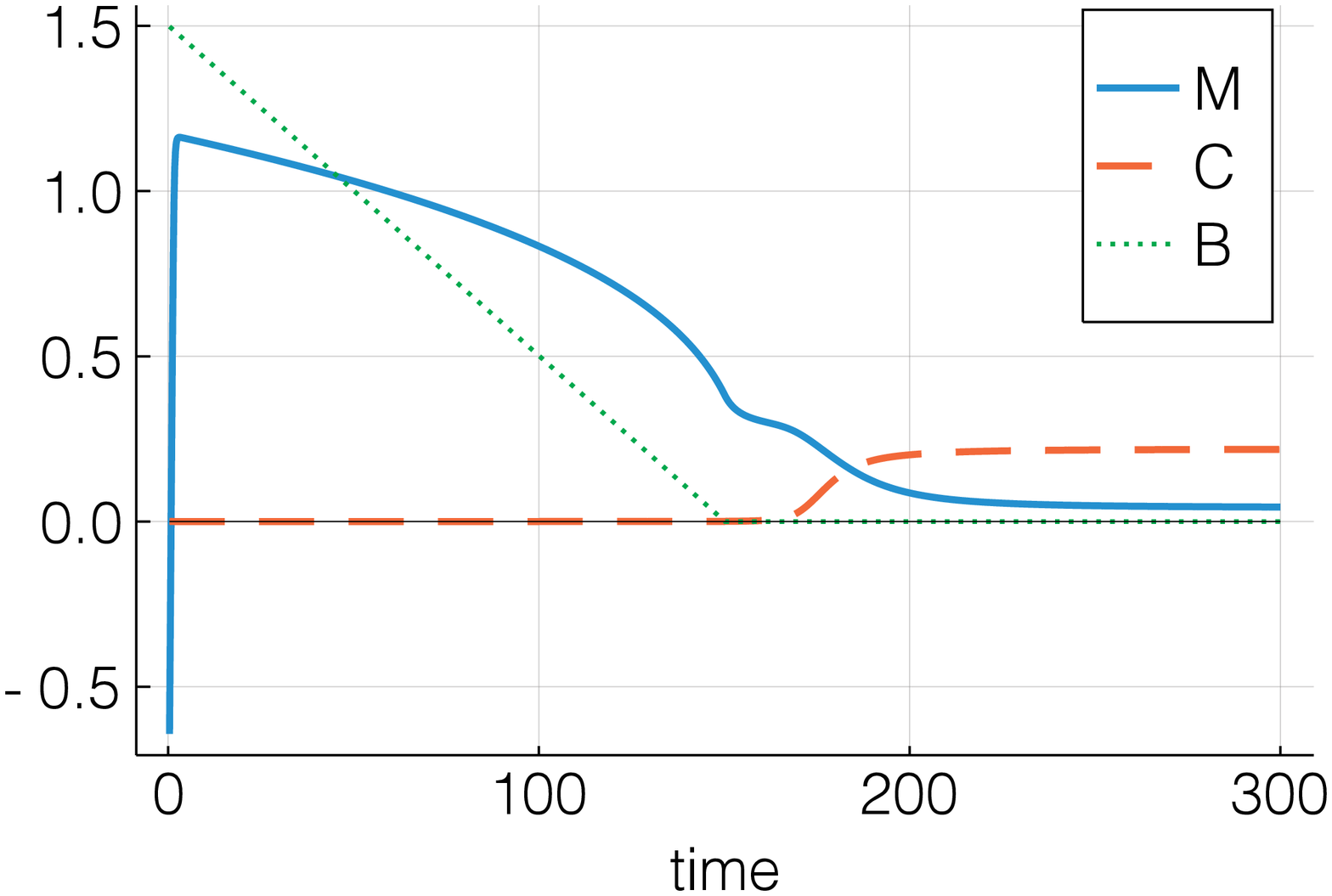}

\vspace{-0.4cm}
\begin{flushleft}
\hspace{0.05\linewidth} (d) Z-breaking phase ($A = 3.0$)
\end{flushleft}
\vspace{-0.3cm}
\includegraphics[width=0.45\linewidth]{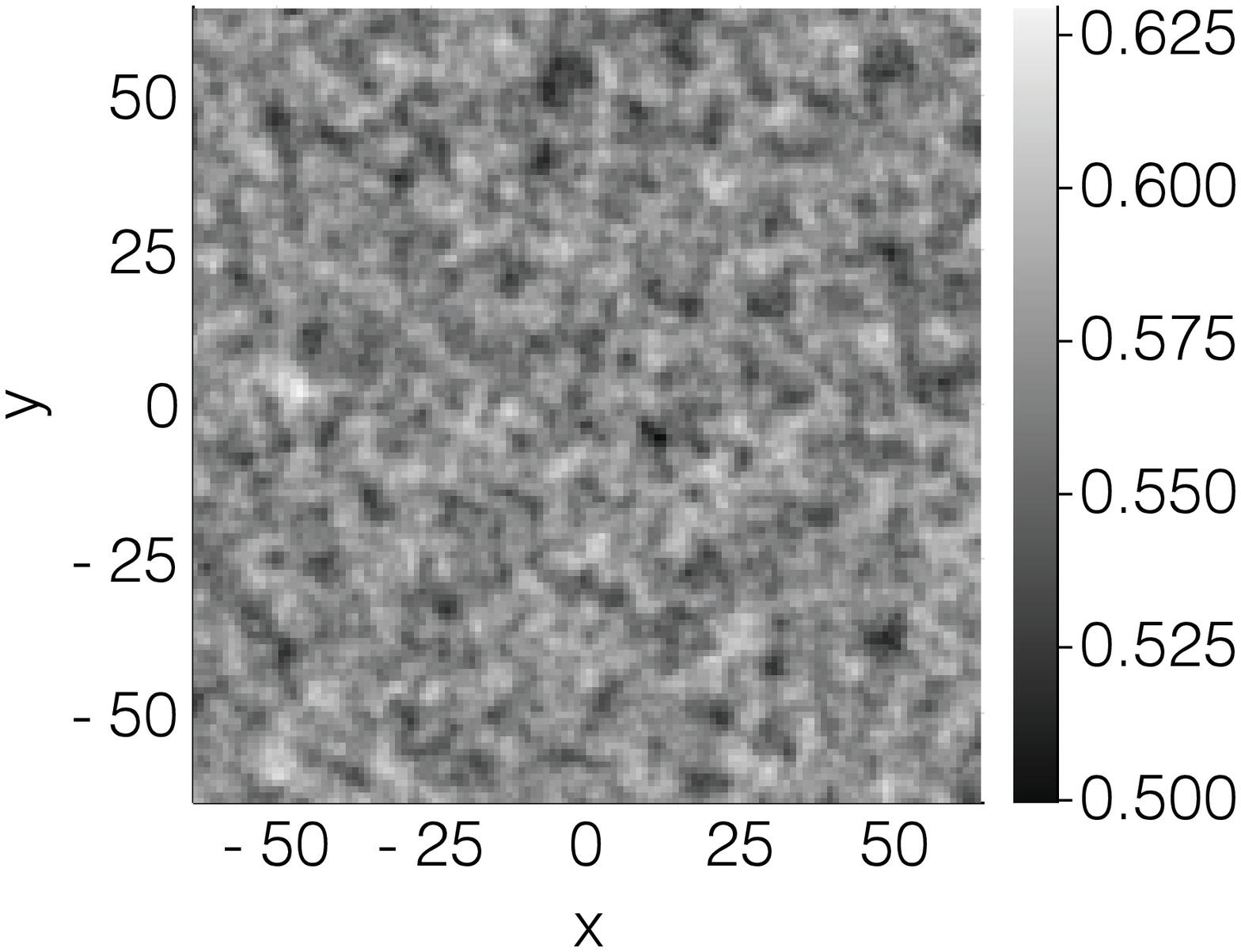}
\includegraphics[width=0.45\linewidth]{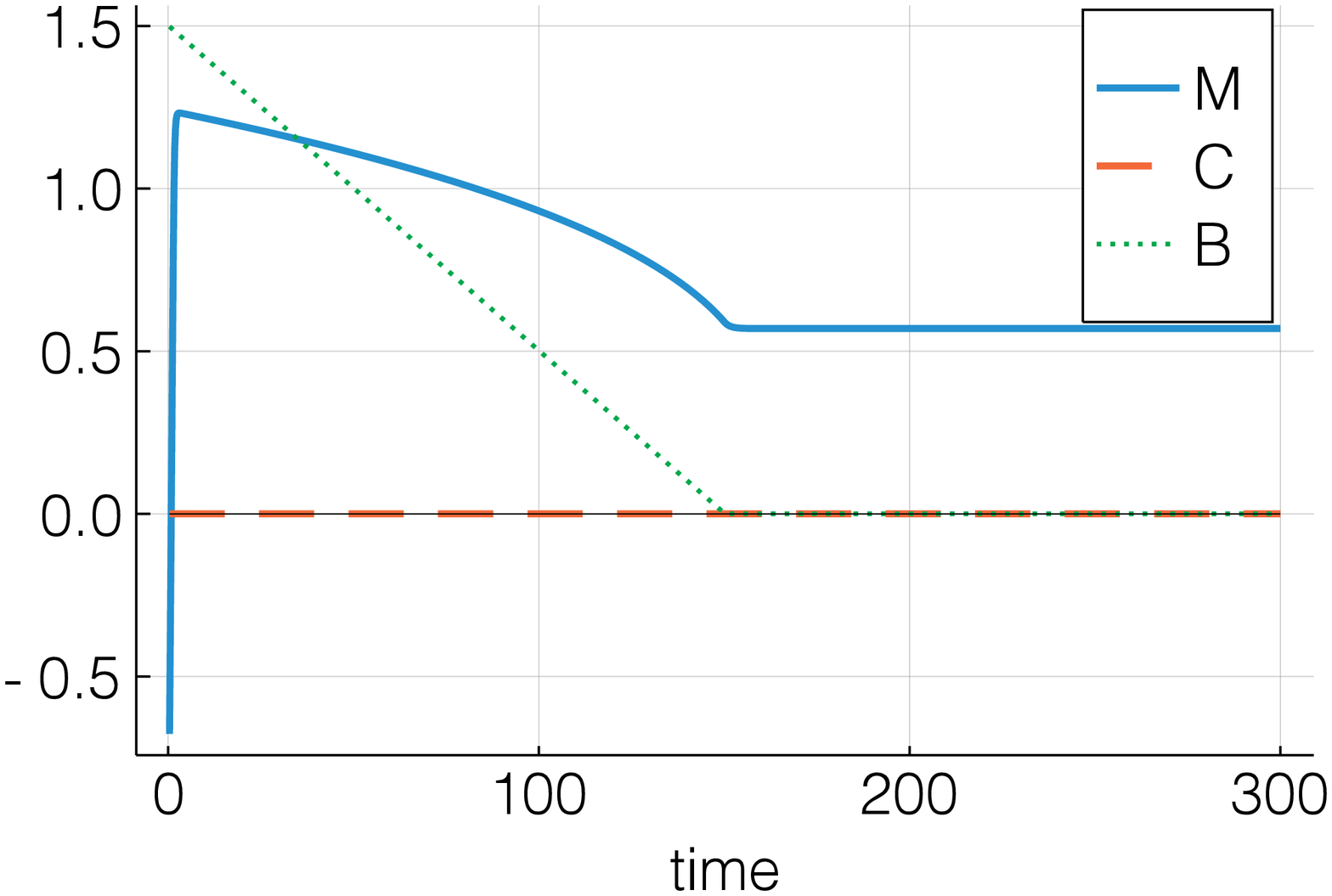}
\vspace{-0.4cm}
\caption{Left Panels: The magnetic domain patterns at the end of the simulation for various $A$. Right Panels: The time dependencies of the external magnetic field (marked by B), the average magnetization $\ev{\phi}_{0}$ (marked by M), and the correlation $\ev{\delta\phi^2}_0$ (marked by C). Simulation size: $L = 512$, duration of the time-evolution: $t_{\max} = 300$, the external magnetic field intensity: $B(t) = R(B_0 - v_{\B}t)$ with $B_0 = 1.5$, $v_{\B} = 0.01$ and $R(x) = \max(x,0)$. The TDSE parameters are: $\alpha = \beta = 1, \gamma = 0.2$. The white regions has positive magnetization, while the black region has negative magnetization.}
\label{fig:Numerical:magnetic}
\end{figure}

The initial spin configuration is prepared in the real-space by distributing $\phi(\rr)$ randomly in a range $-1.1\leq \phi(\rr) \leq -1$ using the uniform distribution. The equation of motion is realized in the wavenumber space so that we can achieve low computational costs for a larger system using the FFT. The resulting spin configuration in the next step is then moved back to the real-space using the inverse FFT (IFFT). The computationally heavy tasks, including the convolutions of the modes $\phi_{\kk}$ in the cubic term, are now circumvented by this method simply by performing the algebraic operation $\phi(\rr)\mapsto[\phi(\rr)]^3$ for each $\rr\in\mathcal{D}$. The time-evolutions are performed efficiently with the ETD2/RK4 method \cite{krogstad2005generalized}, one of the multi-step exponential integrator methods, with relatively large time step $\delta t = 0.2$. The scalability to the system size $L^{2}$ is quite good, with the computational time roughly proportional to $L^{2}$, up to the largest case considered here ($L = 512$). In Fig.~\ref{fig:Numerical:magnetic}, one can see qualitatively different final magnetic domain patterns depending on different values of the thickness of $A$. 

In this system we can define the two-dimensional translational- and rotational symmetry in the coordinate space and the $\mathbf{Z}_{2}$ symmetry of the spin. Thus the patterns above can be naturally classified into four \textit{phases} according to these symmetries: Symmetric [Fig.~\ref{fig:Numerical:magnetic} (a)], T-breaking [(b)], TZ-breaking [(c)], and Z-breaking [(d)] phases, with ``T'' standing for ``translational and rotational'' while ``Z'' standing for $\mathbf{Z}_{2}$. Literatures e.g. \cite{muratov2002theory, jagra2004numerical, kudo2007field} use more descriptive terms referring (b) and (c), such as ``labyrinth'' and ``sea-island'', respectively.

\section{Normalization of the TDGL dynamics}
The physical- or dimension-full TDGL equation Eq.~(\ref{eq:Model:EOM}) is to be normalized by the linear temporal- and spatial coordinate transformations to compare with other results. One of the most convenient choices is to eliminate the dimensionfull saturation magnetization $\rho$. In this case, using the new time variable $\tau$, spatial coordinate $\zzt$, the laplacian $\del^2$ and magnetization $\varphi$, one finds the normalized TDGL equation for the dynamics under the external magnetic field swept from $B_{0} > 0$ to zero with a constant sweep rate $v_{\B}$ is,
\beq\label{eq:normalization:TDGL}
\dv{\varphi}{\tau} = \varphi - \varphi^3 + \del^2\varphi - p_1 F[\varphi] - R(p_2 - p_3\tau),\\
F[\varphi] = \int\de^{2} \zeta' G(\zzt-\zzt')\varphi(\zzt').
\eeq
Here the linear functional $F$ denotes the dipole-dipole interaction, while the coefficients $p_i~(i=1,2,3,4)$ are defined as,
\beq\nonumber
p_1 = \frac{\gamma}{\sqrt{\alpha\beta}}, \quad p_2 = \frac{B_0}{\sqrt{\rho\alpha}}, \quad p_3 = \frac{v_{\B}}{\rho\alpha^2}, \quad p_{4} = \sqrt{\frac{\alpha}{\beta}}A.\\
\eeq
The last parameter $p_{4}$ represents the normalized thickness and is used to construct the effective two-dimensional Green's function as in Eq.~(\ref{Eq:Model:Green}).

\section{equation of balance and RPSA}
If the spin configuration $\phi$ is in the equilibrium state, $\dot{\phi}_{\kk} = 0$ for all $\kk$, and the generic (whether it is normalized or not) equation of motion [Eqs.~(\ref{eq:Model:EOM},\ref{eq:Model:Force})] simplifies into a set of simultaneous time-independent equations. We now introduce a new idea \textit{equation of balance} (EOB), that is the equation of motion in the equilibrium state with zero external magnetization, as shown below.
\beq\label{eq:Balance:b}
\ev{\phi^3}_{\kk} = Q_{\kk}\ev{\phi}_{\kk}; \quad Q_{\kk} = \frac{\alpha -\beta |\kk|^2 - \gamma L^2G_{\kk}}{\alpha}.
\eeq
Let us rewrite Eq.~(\ref{eq:Balance:b}) in the average magnetization $\ev{\phi}_0$ and the modes $\ev{\delta\phi}_{\kk}$ with $\delta\phi(\rr) = \phi(\rr) - \ev{\phi}_0$. By noting that $\ev{\delta\phi}_0 = 0$ and $\ev{c}_{\kk} = 0$ for any constant $c$ and $\kk\neq0$, we immediately obtain the relations governing the balance between the first-, second-, and the third-order moments of the field variables at the equilibrium state. For $\kk = 0$, 
\beq\label{eq:Balance:b0mod}
\left[Q_0 - 3\ev{\delta\phi^{2}}_0\right]\ev{\phi}_{0} = \ev{\delta\phi^3}_{0} + \ev{\phi}_0^3,
\eeq
and for $\kk \neq 0$,
\beq\label{eq:Balance:bneq0mod}
D_{\kk}\ev{\delta\phi}_{\kk} = \ev{\delta\phi^3}_{\kk} + 3\ev{\delta\phi^2}_{\kk}\ev{\phi}_0,
\eeq
with
\beq\label{eq:Balance:bcoeff}
D_{\kk} = Q_{\kk} - 3\ev{\phi}_0^2.
\eeq
These equations do not specify the equilibrium state uniquely. This lack of uniqueness is obvious if one notes that the entire dynamics led to the equilibrium state is not included in the EOB. Thus the EOB must be understood as the restrictions that an equilibrium state must satisfy. 

Since the third-order moment in the EOB can hardly be estimated, we apply the restricted phase-space approximation (RPSA) \cite{anzaki2015restricted} to the equation above. 
In our current context, it is equivalent to a replacement
\beq
\ev{\delta\phi^3}_{\kk} & \to & 3\ev{\delta\phi^2}_0\ev{\delta\phi}_{\kk} \quad (\kk\neq 0).
\eeq
In general, the RPSA truncates the interaction terms $\ev{\delta\phi^3}_{\kk}$ systematically, and known to be exact in special models, e.g $O(N)$ scalar model with $N\to\infty$. 

In the diagrammatic notation, the RPSA is a restriction of the convolution in the Ginzburg-Landau pseudo free energy [corresponding to the equation of motion Eq.(\ref{eq:Model:RHS})] as shown below.

\begin{figure}
\vspace{80pt}
\begin{flushright}
\includegraphics[width = 200pt]{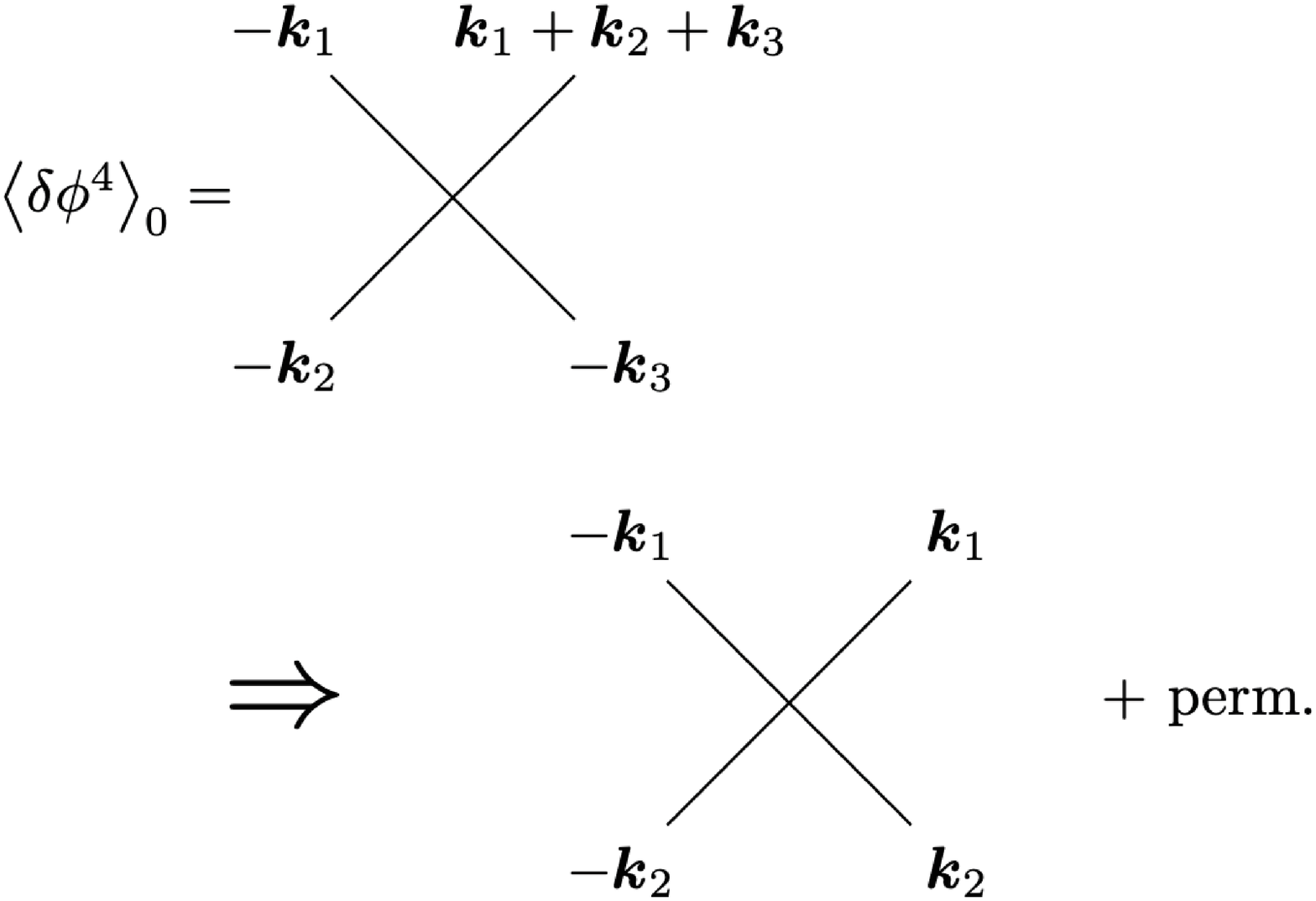}
\end{flushright}
\vspace{-60pt}
\beq
~~
\eeq
\end{figure}
In the diagram above, ``perm.'' indicates the permutations of the vertices. Noting that $\ev{\delta\phi}_0 = 0$, it must be emphasized that the RPSA approximates $\ev{\delta\phi^3}_{0}$ by 0. Thus we obtain the RPSA-EOB as shown below. For $\kk=0$,
\beq\label{eq:Balance:RPSAb0}
\left[Q_0 - 3\ev{\delta\phi^{2}}_0\right]\ev{\phi}_0 = \ev{\phi}_0^3.
\eeq
For $\kk\neq0$,
\beq\label{eq:Balance:RPSAb1}
\left[Q_{\kk} - 3\ev{\phi}^{2}_0 - 3\ev{\delta\phi^{2}}_0\right]\ev{\delta\phi}_{\kk} = 3\ev{\delta\phi^2}_{\kk}\ev{\phi}_0.
\eeq
We confirm that the RPSA agrees with the results of the time-evolutions in the relatively small ($L = 128$) system with $\alpha = 3.5, \beta = 2.0$ and $\gamma = 2/\pi$
. The average magnetization $\ev{\phi}_0$ obtained via the RPSA equation of balance 
using the numerical results of $Q_0$ and $\ev{\delta\phi^2}_0$ matches with the simulation, except for the range $1<A<2.2$, where $\ev{\delta\phi^3}_0$ has nonzero values. This is due to the fact that the RPSA neglects the third-order moment $\ev{\delta\phi^3}_0$ of the distribution.

\section{Phase-prediction method by the RPSA-equation of balance}
\label{Section:methods}
The RPSA-EOB described in the previous Section is applicable for phase predictions of the TDGL dynamics. We simply predict the types of the patterns based on the nonzero modes under the restrictions imposed by the RPSA-EOB [Eqs.~(\ref{eq:Balance:RPSAb0},\ref{eq:Balance:RPSAb1})]. We use the continuum EOM [Eq.~\ref{eq:Model:continuum}] to use the analytic form of the Green's function [Eq.~(\ref{eq:Model:Green:analytic})]. Note that this choice causes a modification to the definition of $Q_{\kk}$:
\beq
Q_{\kk} = \frac{\alpha -\beta |\kk|^2 - \gamma (2\pi)^2G_{\kk}(A)}{\alpha},
\eeq
with $G_{\kk}(A)$ being the continuum limit of the Green's function shown in Eq.~(\ref{eq:Model:Green:analytic}). The method \texttt{PhasePrediction} is schematically shown below. This is a procedure that maps a set of parameters $(\alpha,\beta,\gamma,A)$ to the output \texttt{Phase} $\in \{$\texttt{Symmetric, T-breaking, *Z-breaking}$\}$, with \texttt{*Z-breaking} means either TZ-breaking or Z-breaking. 

\vspace{12pt}
 \begin{algorithmic}[1]
  \Procedure{PhasePrediction}{$\alpha,\beta,\gamma,A$}
  \State{$g(k) = (1-\e^{-Ak})/(\pi A^2k)$}
  \State{$Q(k) = 1 - \alpha^{-1}\beta k^2 - \alpha^{-1}\gamma (2\pi)^2g(k)$}
 			\If{$Q(0) < 0$}
   	\If{$\max(Q) < 0$}
   		\State{\texttt{Phase} $\Leftarrow$ \texttt{Symmetric}}
   	\Else
   		\State{\texttt{Phase} $\Leftarrow$ \texttt{T-breaking}}
   	\EndIf
			\Else
					\State{\texttt{Phase} $\Leftarrow$ \texttt{*Z-breaking}}
   \EndIf

  \EndProcedure
 \end{algorithmic}
\vspace{12pt}

\section{Discussion}
\begin{figure}
\includegraphics[width=0.9\linewidth]{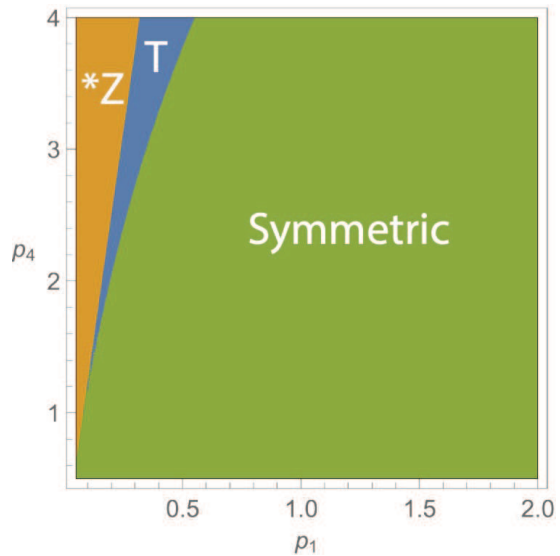}
\caption{The phase diagram of the TDGL dynamics corresponding to Eq.~(\ref{eq:normalization:TDGL}) estimated by the method \texttt{PhasePrediction} proposed in Sec. \ref{Section:methods}. Orange area (marked with ``*Z''): TZ- or Z-breaking phase, blue area (``T''): T-breaking phase, green area (``Symmetric''): symmetric phase.}
\label{fig:phase:phase-diagram-analytic}
\end{figure}

The phase diagram for the normalized TDGL dynamics [Eq.~(\ref{eq:normalization:TDGL})] predicted by the method \texttt{PhasePrediction} described in Sec. \ref{Section:methods} is shown in Fig.~\ref{fig:phase:phase-diagram-analytic}. The overall tendency matches our physical instinct well. As $p_1$ becomes large, the demagnetization effect from the dipole-dipole interactions supersedes the anisotropy to yield the symmetric phase, while for larger $p_4$, it is partly relaxed by the thickness to have more complexed structures. 

We also compared the numerical results of time-evolution with the phase prediction. The results are shown in Table~\ref{table:Phase:comparison}. The computational cost for each sample point $(p_1,p_4)$ is significantly small compared to the corresponding numerical time-evolutions. 

The agreement between the time-evolution and the phase prediction is good, except in the cases $(p_1,p_4) = (0.1,1.0), (0.4,2.5), (0.4,3.0)$. This is due to the relatively small absolute values of $\max(Q)$ and $Q(0)$ at these sample points. Since the RPSA neglects the third-order moments in the EOBs, the results obtained by the RPSA-EOB based method may differ from the time-evolution for small $|Q(0)|$ and $|\max(Q)|$. This mismatch may improve by further developments of the approximation; in other words, it is considered that the third-order moments play crucial roles in the region where the mismatch is seen.



\begin{table}
\caption{Comparison with numerical time-evolutions with \texttt{PhasePrediction}. T, Z, TZ, and S denotes the phases observed from the time-evolution (T-, Z-, TZ-breaking, and symmetric phases, respectively), while $(s,t)$ with $s,t\in\{+,-\}$ denotes the signs of the $Q_0$ and $\max(Q)$, respectively. Note that \texttt{PhasePrediction} translates $(++)$ to *Z ($=$ Z, TZ), $(-+)$ to T, and $(--)$ to S. The numerical time-evolutions are performed for the system with $512^2$ grid points using the normalized TDGL equation Eq.(\ref{eq:normalization:TDGL}). Points marked with ``$\star$'' indicate the mismatches with the numerical results.}
\vspace{4pt}
\begin{tabular}{c|c|c|c|c|c}
$p_4 \backslash p_1$ & 0.1 & 0.2 & 0.4 & 0.8 & 1.6 \\ \hline
1.0 & Z$^{\star}$($--$) & S($--$) & S($--$) & S($--$) & S($--$) \\
1.5 & Z($++$) & T($--$) & S($--$) & S($--$) & S($--$) \\
2.0 & Z($++$) & TZ($-+$) & S($--$) & S($--$) & S($--$) \\
2.5 & Z($++$) & Z($++$) & T$^{\star}$($--$) & S($--$) & S($--$) \\
3.0 & Z($++$) & Z($++$) & T$^{\star}$($--$) & S($--$) & S($--$) \\
\end{tabular}
\label{table:Phase:comparison}
\end{table}

Note that the external magnetic sweep rate $v_{\B}$ is an important parameter in the pattern formation. It is reported \cite{kudo2007field} that the domain formation is largely affected by $v_{\B}$. Our results here must be understood as an approximated result, not only in the RPSA but also in the elimination of the effects of the magnetic sweep rate. In the $v_{\B}\to\infty$ limit, our method will have the same results, while that of the time-evolutions can be quite different.

Although this method does not have a perfect concordance with the numerical simulations, it has no \textit{a priori} parameters or functions in any form, but only approximated in a systematic, physically reasonable way \cite{anzaki2015restricted}. Hence it is considered as a method without any \textit{a priori} assumptions on the domain patterns. This fact means that one can \textit{add} new features, e.g., tuning parameters, without doubting the physical meaning of this method, provided the approximation is reasonable.  

\section{Conclusions}
The long history of the research in the magnetism and the mathematical structure of the TDGL dynamics show a wide variety of approaches to the pattern formation in the magnetic materials \cite{kittel1946theory, kooy1960, bochi1995magnetic, kaplan1993domain, lisfi2002magnetic}. 

Although most of the existing methods use artificial functions that specify the magnetic domain patterns, we focus on the \textit{equation of balance} (EOB) that a magnetic material must satisfy in its equilibrium state. Applying the restricted phase-space approximation (RPSA) \cite{anzaki2015restricted} to EOB enables us to predict the phase in the equilibrium state. The prediction matches the actual numerical time-evolution results qualitatively without any tuning parameters. Although the prediction is not perfect, our method has no \textit{a priori} assumptions, i.e., it does not involve any artificial function or experimentally justified parameters but only approximated systematically. Thus it is very extensive, applicable for various applications.  

Another aspect that must be noted is that the object of the new method is not limited to the magnetic systems; it is applicable for a vast class of natural/social phenomena that seemingly have nothing in common but described by the equation of motion of type Eqs.(\ref{eq:Model:EOM},\ref{eq:Model:Force}).


One of such applications is parameter estimation in material- and statistical physics. Using the Bayesian inference methods, we can estimate the parameters of a system with huge degrees of freedom by relatively small observation/numerical data, e.g., \cite{ito2019bayesian}. Our method is expected to serve for such parameter estimations in various systems as the theoretical- and numerical basis by giving information on the phase for each parameter using a few computational costs, with physically justifiable reasons.
\vspace{12pt}

\section*{Acknowledgements}
This work was mainly supported by JST CREST Grant Numbers JPMJCR1761 and JPMJCR1861 and partially supported by JPMJCR1763 of Japan Science and Technology Agency. 
The key ideas in this study came through the activities of JSPS KAKENHI 
Grant Numbers JP19K14671, 
JP17H01703, JP17H01704, JP18H03210, 
JP19H05662, 
and
JP20K21785. 
The travel expense needed to discuss among co-authors was partially supported by ERI JURP 2020-A-05, 2018-B-01, and 2019-B-04. 

%
%

\vspace{\fill}

\bibliography{ms}

\begin{thebibliography}{30}%
\makeatletter
\providecommand \@ifxundefined [1]{%
 \@ifx{#1\undefined}
}%
\providecommand \@ifnum [1]{%
 \ifnum #1\expandafter \@firstoftwo
 \else \expandafter \@secondoftwo
 \fi
}%
\providecommand \@ifx [1]{%
 \ifx #1\expandafter \@firstoftwo
 \else \expandafter \@secondoftwo
 \fi
}%
\providecommand \natexlab [1]{#1}%
\providecommand \enquote  [1]{``#1''}%
\providecommand \bibnamefont  [1]{#1}%
\providecommand \bibfnamefont [1]{#1}%
\providecommand \citenamefont [1]{#1}%
\providecommand \href@noop [0]{\@secondoftwo}%
\providecommand \href [0]{\begingroup \@sanitize@url \@href}%
\providecommand \@href[1]{\@@startlink{#1}\@@href}%
\providecommand \@@href[1]{\endgroup#1\@@endlink}%
\providecommand \@sanitize@url [0]{\catcode `\\12\catcode `\$12\catcode
  `\&12\catcode `\#12\catcode `\^12\catcode `\_12\catcode `\%12\relax}%
\providecommand \@@startlink[1]{}%
\providecommand \@@endlink[0]{}%
\providecommand \url  [0]{\begingroup\@sanitize@url \@url }%
\providecommand \@url [1]{\endgroup\@href {#1}{\urlprefix }}%
\providecommand \urlprefix  [0]{URL }%
\providecommand \Eprint [0]{\href }%
\providecommand \doibase [0]{https://doi.org/}%
\providecommand \selectlanguage [0]{\@gobble}%
\providecommand \bibinfo  [0]{\@secondoftwo}%
\providecommand \bibfield  [0]{\@secondoftwo}%
\providecommand \translation [1]{[#1]}%
\providecommand \BibitemOpen [0]{}%
\providecommand \bibitemStop [0]{}%
\providecommand \bibitemNoStop [0]{.\EOS\space}%
\providecommand \EOS [0]{\spacefactor3000\relax}%
\providecommand \BibitemShut  [1]{\csname bibitem#1\endcsname}%
\let\auto@bib@innerbib\@empty
\bibitem [{\citenamefont {Van~Vleck}(1945)}]{vleck1945survey}%
  \BibitemOpen
  \bibfield  {author} {\bibinfo {author} {\bibfnamefont {J.~H.}\ \bibnamefont
  {Van~Vleck}},\ }\href {https://doi.org/10.1103/RevModPhys.17.27} {\bibfield
  {journal} {\bibinfo  {journal} {Rev. Mod. Phys.}\ }\textbf {\bibinfo {volume}
  {17}},\ \bibinfo {pages} {27} (\bibinfo {year} {1945})}\BibitemShut {NoStop}%
\bibitem [{\citenamefont {Kittel}(1949)}]{kittel1949physical}%
  \BibitemOpen
  \bibfield  {author} {\bibinfo {author} {\bibfnamefont {C.}~\bibnamefont
  {Kittel}},\ }\href {https://doi.org/10.1103/RevModPhys.21.541} {\bibfield
  {journal} {\bibinfo  {journal} {Rev. Mod. Phys.}\ }\textbf {\bibinfo {volume}
  {21}},\ \bibinfo {pages} {541} (\bibinfo {year} {1949})}\BibitemShut
  {NoStop}%
\bibitem [{\citenamefont {Suzuki}\ \emph {et~al.}(2013)\citenamefont {Suzuki},
  \citenamefont {Kawamura}, \citenamefont {Mizumaki}, \citenamefont {Terada},
  \citenamefont {Uruga}, \citenamefont {Fujiwara}, \citenamefont {Yamazaki},
  \citenamefont {Yumoto}, \citenamefont {Koyama}, \citenamefont {Senba} \emph
  {et~al.}}]{suzuki2013hard}%
  \BibitemOpen
  \bibfield  {author} {\bibinfo {author} {\bibfnamefont {M.}~\bibnamefont
  {Suzuki}}, \bibinfo {author} {\bibfnamefont {N.}~\bibnamefont {Kawamura}},
  \bibinfo {author} {\bibfnamefont {M.}~\bibnamefont {Mizumaki}}, \bibinfo
  {author} {\bibfnamefont {Y.}~\bibnamefont {Terada}}, \bibinfo {author}
  {\bibfnamefont {T.}~\bibnamefont {Uruga}}, \bibinfo {author} {\bibfnamefont
  {A.}~\bibnamefont {Fujiwara}}, \bibinfo {author} {\bibfnamefont
  {H.}~\bibnamefont {Yamazaki}}, \bibinfo {author} {\bibfnamefont
  {H.}~\bibnamefont {Yumoto}}, \bibinfo {author} {\bibfnamefont
  {T.}~\bibnamefont {Koyama}}, \bibinfo {author} {\bibfnamefont
  {Y.}~\bibnamefont {Senba}}, \emph {et~al.},\ }in\ \href@noop {} {\emph
  {\bibinfo {booktitle} {J. Phys. Conf. Ser.}}},\ Vol.\ \bibinfo {volume}
  {430}\ (\bibinfo {year} {2013})\BibitemShut {NoStop}%
\bibitem [{\citenamefont {Argyres}(1955)}]{argyres1955theory}%
  \BibitemOpen
  \bibfield  {author} {\bibinfo {author} {\bibfnamefont {P.~N.}\ \bibnamefont
  {Argyres}},\ }\href {https://doi.org/10.1103/PhysRev.97.334} {\bibfield
  {journal} {\bibinfo  {journal} {Phys. Rev.}\ }\textbf {\bibinfo {volume}
  {97}},\ \bibinfo {pages} {334} (\bibinfo {year} {1955})}\BibitemShut
  {NoStop}%
\bibitem [{\citenamefont {Reif}\ \emph {et~al.}(1991)\citenamefont {Reif},
  \citenamefont {Zink}, \citenamefont {Schneider},\ and\ \citenamefont
  {Kirschner}}]{reif1991effects}%
  \BibitemOpen
  \bibfield  {author} {\bibinfo {author} {\bibfnamefont {J.}~\bibnamefont
  {Reif}}, \bibinfo {author} {\bibfnamefont {J.~C.}\ \bibnamefont {Zink}},
  \bibinfo {author} {\bibfnamefont {C.-M.}\ \bibnamefont {Schneider}},\ and\
  \bibinfo {author} {\bibfnamefont {J.}~\bibnamefont {Kirschner}},\ }\href
  {https://doi.org/10.1103/PhysRevLett.67.2878} {\bibfield  {journal} {\bibinfo
   {journal} {Phys. Rev. Lett.}\ }\textbf {\bibinfo {volume} {67}},\ \bibinfo
  {pages} {2878} (\bibinfo {year} {1991})}\BibitemShut {NoStop}%
\bibitem [{\citenamefont {Jagla}(2004)}]{jagra2004numerical}%
  \BibitemOpen
  \bibfield  {author} {\bibinfo {author} {\bibfnamefont {E.~A.}\ \bibnamefont
  {Jagla}},\ }\href {https://doi.org/10.1103/PhysRevE.70.046204} {\bibfield
  {journal} {\bibinfo  {journal} {Phys. Rev. E}\ }\textbf {\bibinfo {volume}
  {70}},\ \bibinfo {pages} {046204} (\bibinfo {year} {2004})}\BibitemShut
  {NoStop}%
\bibitem [{\citenamefont {Kudo}\ and\ \citenamefont
  {Nakamura}(2007)}]{kudo2007field}%
  \BibitemOpen
  \bibfield  {author} {\bibinfo {author} {\bibfnamefont {K.}~\bibnamefont
  {Kudo}}\ and\ \bibinfo {author} {\bibfnamefont {K.}~\bibnamefont
  {Nakamura}},\ }\href {https://doi.org/10.1103/PhysRevB.76.054111} {\bibfield
  {journal} {\bibinfo  {journal} {Phys. Rev. B}\ }\textbf {\bibinfo {volume}
  {76}},\ \bibinfo {pages} {054111} (\bibinfo {year} {2007})}\BibitemShut
  {NoStop}%
\bibitem [{\citenamefont {Iwano}\ \emph {et~al.}(2014)\citenamefont {Iwano},
  \citenamefont {Mitsumata},\ and\ \citenamefont {Ono}}]{iwano2014maze}%
  \BibitemOpen
  \bibfield  {author} {\bibinfo {author} {\bibfnamefont {K.}~\bibnamefont
  {Iwano}}, \bibinfo {author} {\bibfnamefont {C.}~\bibnamefont {Mitsumata}},\
  and\ \bibinfo {author} {\bibfnamefont {K.}~\bibnamefont {Ono}},\ }\href
  {https://doi.org/10.1063/1.4865779} {\bibfield  {journal} {\bibinfo
  {journal} {J. Appl. Phys.}\ }\textbf {\bibinfo {volume} {115}},\ \bibinfo
  {pages} {17D134} (\bibinfo {year} {2014})}\BibitemShut {NoStop}%
\bibitem [{\citenamefont
  {Kawasaki}(1974{\natexlab{a}})}]{kawasaki1974macroscopic}%
  \BibitemOpen
  \bibfield  {author} {\bibinfo {author} {\bibfnamefont {K.}~\bibnamefont
  {Kawasaki}},\ }\href@noop {} {\bibfield  {journal} {\bibinfo  {journal}
  {Prog. Theor. Phys.}\ }\textbf {\bibinfo {volume} {52}},\ \bibinfo {pages}
  {359} (\bibinfo {year} {1974}{\natexlab{a}})}\BibitemShut {NoStop}%
\bibitem [{\citenamefont
  {Kawasaki}(1974{\natexlab{b}})}]{kawasaki1974contributions}%
  \BibitemOpen
  \bibfield  {author} {\bibinfo {author} {\bibfnamefont {K.}~\bibnamefont
  {Kawasaki}},\ }\href {https://doi.org/10.1143/PTP.51.1064} {\bibfield
  {journal} {\bibinfo  {journal} {Prog. Theor. Phys.}\ }\textbf {\bibinfo
  {volume} {51}},\ \bibinfo {pages} {1064} (\bibinfo {year}
  {1974}{\natexlab{b}})}\BibitemShut {NoStop}%
\bibitem [{\citenamefont
  {Kawasaki}(1974{\natexlab{c}})}]{kawasaki1974contribution}%
  \BibitemOpen
  \bibfield  {author} {\bibinfo {author} {\bibfnamefont {K.}~\bibnamefont
  {Kawasaki}},\ }\href {https://doi.org/10.1143/PTP.52.84} {\bibfield
  {journal} {\bibinfo  {journal} {Prog. Theor. Phys.}\ }\textbf {\bibinfo
  {volume} {52}},\ \bibinfo {pages} {84} (\bibinfo {year}
  {1974}{\natexlab{c}})}\BibitemShut {NoStop}%
\bibitem [{\citenamefont {Suzuki}\ and\ \citenamefont
  {Igarashi}(1973)}]{suzuki1973calculation}%
  \BibitemOpen
  \bibfield  {author} {\bibinfo {author} {\bibfnamefont {M.}~\bibnamefont
  {Suzuki}}\ and\ \bibinfo {author} {\bibfnamefont {G.}~\bibnamefont
  {Igarashi}},\ }\href@noop {} {\bibfield  {journal} {\bibinfo  {journal}
  {Prog. Theor. Phys.}\ }\textbf {\bibinfo {volume} {49}},\ \bibinfo {pages}
  {1070} (\bibinfo {year} {1973})}\BibitemShut {NoStop}%
\bibitem [{\citenamefont {Grant}\ \emph {et~al.}(1985)\citenamefont {Grant},
  \citenamefont {San~Miguel}, \citenamefont {Vials},\ and\ \citenamefont
  {Gunton}}]{grant1985theory}%
  \BibitemOpen
  \bibfield  {author} {\bibinfo {author} {\bibfnamefont {M.}~\bibnamefont
  {Grant}}, \bibinfo {author} {\bibfnamefont {M.}~\bibnamefont {San~Miguel}},
  \bibinfo {author} {\bibfnamefont {J.}~\bibnamefont {Vials}},\ and\ \bibinfo
  {author} {\bibfnamefont {J.~D.}\ \bibnamefont {Gunton}},\ }\href
  {https://doi.org/10.1103/PhysRevB.31.3027} {\bibfield  {journal} {\bibinfo
  {journal} {Phys. Rev. B}\ }\textbf {\bibinfo {volume} {31}},\ \bibinfo
  {pages} {3027} (\bibinfo {year} {1985})}\BibitemShut {NoStop}%
\bibitem [{\citenamefont {Kawasaki}\ and\ \citenamefont
  {Ohta}(1982)}]{kawasaki1982kink}%
  \BibitemOpen
  \bibfield  {author} {\bibinfo {author} {\bibfnamefont {K.}~\bibnamefont
  {Kawasaki}}\ and\ \bibinfo {author} {\bibfnamefont {T.}~\bibnamefont
  {Ohta}},\ }\href@noop {} {\bibfield  {journal} {\bibinfo  {journal} {Physica
  A}\ }\textbf {\bibinfo {volume} {116}},\ \bibinfo {pages} {573} (\bibinfo
  {year} {1982})}\BibitemShut {NoStop}%
\bibitem [{\citenamefont {Yasui}\ \emph {et~al.}(2002)\citenamefont {Yasui},
  \citenamefont {Tutu}, \citenamefont {Yamamoto},\ and\ \citenamefont
  {Fujisaka}}]{yasui2002dynamic}%
  \BibitemOpen
  \bibfield  {author} {\bibinfo {author} {\bibfnamefont {T.}~\bibnamefont
  {Yasui}}, \bibinfo {author} {\bibfnamefont {H.}~\bibnamefont {Tutu}},
  \bibinfo {author} {\bibfnamefont {M.}~\bibnamefont {Yamamoto}},\ and\
  \bibinfo {author} {\bibfnamefont {H.}~\bibnamefont {Fujisaka}},\ }\href
  {https://doi.org/10.1103/PhysRevE.66.036123} {\bibfield  {journal} {\bibinfo
  {journal} {Phys. Rev. E}\ }\textbf {\bibinfo {volume} {66}},\ \bibinfo
  {pages} {036123} (\bibinfo {year} {2002})}\BibitemShut {NoStop}%
\bibitem [{\citenamefont {Fujiwara}\ \emph {et~al.}(2004)\citenamefont
  {Fujiwara}, \citenamefont {Tutu},\ and\ \citenamefont
  {Fujisaka}}]{fujiwara2004magnetic}%
  \BibitemOpen
  \bibfield  {author} {\bibinfo {author} {\bibfnamefont {N.}~\bibnamefont
  {Fujiwara}}, \bibinfo {author} {\bibfnamefont {H.}~\bibnamefont {Tutu}},\
  and\ \bibinfo {author} {\bibfnamefont {H.}~\bibnamefont {Fujisaka}},\ }\href
  {https://doi.org/10.1103/PhysRevE.70.066132} {\bibfield  {journal} {\bibinfo
  {journal} {Phys. Rev. E}\ }\textbf {\bibinfo {volume} {70}},\ \bibinfo
  {pages} {066132} (\bibinfo {year} {2004})}\BibitemShut {NoStop}%
\bibitem [{\citenamefont {Rogers}\ \emph {et~al.}(1988)\citenamefont {Rogers},
  \citenamefont {Elder},\ and\ \citenamefont {Desai}}]{rogers1988numerical}%
  \BibitemOpen
  \bibfield  {author} {\bibinfo {author} {\bibfnamefont {T.~M.}\ \bibnamefont
  {Rogers}}, \bibinfo {author} {\bibfnamefont {K.~R.}\ \bibnamefont {Elder}},\
  and\ \bibinfo {author} {\bibfnamefont {R.~C.}\ \bibnamefont {Desai}},\ }\href
  {https://doi.org/10.1103/PhysRevB.37.9638} {\bibfield  {journal} {\bibinfo
  {journal} {Phys. Rev. B}\ }\textbf {\bibinfo {volume} {37}},\ \bibinfo
  {pages} {9638} (\bibinfo {year} {1988})}\BibitemShut {NoStop}%
\bibitem [{\citenamefont {Kittel}(1946)}]{kittel1946theory}%
  \BibitemOpen
  \bibfield  {author} {\bibinfo {author} {\bibfnamefont {C.}~\bibnamefont
  {Kittel}},\ }\href {https://doi.org/10.1103/PhysRev.70.965} {\bibfield
  {journal} {\bibinfo  {journal} {Phys. Rev.}\ }\textbf {\bibinfo {volume}
  {70}},\ \bibinfo {pages} {965} (\bibinfo {year} {1946})}\BibitemShut
  {NoStop}%
\bibitem [{\citenamefont {Kaplan}\ and\ \citenamefont
  {Gehring}(1993)}]{kaplan1993domain}%
  \BibitemOpen
  \bibfield  {author} {\bibinfo {author} {\bibfnamefont {B.}~\bibnamefont
  {Kaplan}}\ and\ \bibinfo {author} {\bibfnamefont {G.}~\bibnamefont
  {Gehring}},\ }\href
  {https://doi.org/https://doi.org/10.1016/0304-8853(93)90863-W} {\bibfield
  {journal} {\bibinfo  {journal} {J. Magn. Magn. Mater}\ }\textbf {\bibinfo
  {volume} {128}},\ \bibinfo {pages} {111 } (\bibinfo {year}
  {1993})}\BibitemShut {NoStop}%
\bibitem [{\citenamefont {Bochi}\ \emph {et~al.}(1995)\citenamefont {Bochi},
  \citenamefont {Hug}, \citenamefont {Paul}, \citenamefont {Stiefel},
  \citenamefont {Moser}, \citenamefont {Parashikov}, \citenamefont
  {G\"untherodt},\ and\ \citenamefont {O'Handley}}]{bochi1995magnetic}%
  \BibitemOpen
  \bibfield  {author} {\bibinfo {author} {\bibfnamefont {G.}~\bibnamefont
  {Bochi}}, \bibinfo {author} {\bibfnamefont {H.~J.}\ \bibnamefont {Hug}},
  \bibinfo {author} {\bibfnamefont {D.~I.}\ \bibnamefont {Paul}}, \bibinfo
  {author} {\bibfnamefont {B.}~\bibnamefont {Stiefel}}, \bibinfo {author}
  {\bibfnamefont {A.}~\bibnamefont {Moser}}, \bibinfo {author} {\bibfnamefont
  {I.}~\bibnamefont {Parashikov}}, \bibinfo {author} {\bibfnamefont {H.-J.}\
  \bibnamefont {G\"untherodt}},\ and\ \bibinfo {author} {\bibfnamefont {R.~C.}\
  \bibnamefont {O'Handley}},\ }\href
  {https://doi.org/10.1103/PhysRevLett.75.1839} {\bibfield  {journal} {\bibinfo
   {journal} {Phys. Rev. Lett.}\ }\textbf {\bibinfo {volume} {75}},\ \bibinfo
  {pages} {1839} (\bibinfo {year} {1995})}\BibitemShut {NoStop}%
\bibitem [{\citenamefont {Kooy}\ and\ \citenamefont {Enz}(1960)}]{kooy1960}%
  \BibitemOpen
  \bibfield  {author} {\bibinfo {author} {\bibfnamefont {C.}~\bibnamefont
  {Kooy}}\ and\ \bibinfo {author} {\bibfnamefont {U.}~\bibnamefont {Enz}},\
  }\href@noop {} {\bibfield  {journal} {\bibinfo  {journal} {Philips Research
  Reports}\ }\textbf {\bibinfo {volume} {15}} (\bibinfo {year}
  {1960})}\BibitemShut {NoStop}%
\bibitem [{\citenamefont {Lisfi}\ and\ \citenamefont
  {Lodder}(2002)}]{lisfi2002magnetic}%
  \BibitemOpen
  \bibfield  {author} {\bibinfo {author} {\bibfnamefont {A.}~\bibnamefont
  {Lisfi}}\ and\ \bibinfo {author} {\bibfnamefont {J.~C.}\ \bibnamefont
  {Lodder}},\ }\href {https://doi.org/10.1088/0953-8984/14/47/309} {\bibfield
  {journal} {\bibinfo  {journal} {J. Phys. Condens. Matter}\ }\textbf {\bibinfo
  {volume} {14}},\ \bibinfo {pages} {12339} (\bibinfo {year}
  {2002})}\BibitemShut {NoStop}%
\bibitem [{\citenamefont {Garel}\ and\ \citenamefont
  {Doniach}(1982)}]{garel1982phase}%
  \BibitemOpen
  \bibfield  {author} {\bibinfo {author} {\bibfnamefont {T.}~\bibnamefont
  {Garel}}\ and\ \bibinfo {author} {\bibfnamefont {S.}~\bibnamefont
  {Doniach}},\ }\href {https://doi.org/10.1103/PhysRevB.26.325} {\bibfield
  {journal} {\bibinfo  {journal} {Phys. Rev. B}\ }\textbf {\bibinfo {volume}
  {26}},\ \bibinfo {pages} {325} (\bibinfo {year} {1982})}\BibitemShut
  {NoStop}%
\bibitem [{\citenamefont {Kudo}\ \emph {et~al.}()\citenamefont {Kudo},
  \citenamefont {Mino},\ and\ \citenamefont {Nakamura}}]{kudo2007magnetic}%
  \BibitemOpen
  \bibfield  {author} {\bibinfo {author} {\bibfnamefont {K.}~\bibnamefont
  {Kudo}}, \bibinfo {author} {\bibfnamefont {M.}~\bibnamefont {Mino}},\ and\
  \bibinfo {author} {\bibfnamefont {K.}~\bibnamefont {Nakamura}},\ }\href@noop
  {} {\bibinfo  {journal} {J. Phys. Soc. Japan}\ }\BibitemShut {NoStop}%
\bibitem [{\citenamefont {Yokota}(2017)}]{yokota2017three}%
  \BibitemOpen
\bibfield  {journal} {  }\bibfield  {author} {\bibinfo {author} {\bibfnamefont
  {T.}~\bibnamefont {Yokota}},\ }\href@noop {} {\bibfield  {journal} {\bibinfo
  {journal} {J. Magn. Magn. Mater.}\ }\textbf {\bibinfo {volume} {432}},\
  \bibinfo {pages} {532} (\bibinfo {year} {2017})}\BibitemShut {NoStop}%
\bibitem [{\citenamefont {Tom\'e}\ and\ \citenamefont
  {de~Oliveira}(1990)}]{tome1990dynamic}%
  \BibitemOpen
  \bibfield  {author} {\bibinfo {author} {\bibfnamefont {T.}~\bibnamefont
  {Tom\'e}}\ and\ \bibinfo {author} {\bibfnamefont {M.~J.}\ \bibnamefont
  {de~Oliveira}},\ }\href {https://doi.org/10.1103/PhysRevA.41.4251} {\bibfield
   {journal} {\bibinfo  {journal} {Phys. Rev. A}\ }\textbf {\bibinfo {volume}
  {41}},\ \bibinfo {pages} {4251} (\bibinfo {year} {1990})}\BibitemShut
  {NoStop}%
\bibitem [{\citenamefont {Krogstad}(2005)}]{krogstad2005generalized}%
  \BibitemOpen
  \bibfield  {author} {\bibinfo {author} {\bibfnamefont {S.}~\bibnamefont
  {Krogstad}},\ }\href@noop {} {\bibfield  {journal} {\bibinfo  {journal} {J.
  Comput. Phys.}\ }\textbf {\bibinfo {volume} {203}},\ \bibinfo {pages} {72}
  (\bibinfo {year} {2005})}\BibitemShut {NoStop}%
\bibitem [{\citenamefont {Muratov}(2002)}]{muratov2002theory}%
  \BibitemOpen
  \bibfield  {author} {\bibinfo {author} {\bibfnamefont {C.~B.}\ \bibnamefont
  {Muratov}},\ }\href {https://doi.org/10.1103/PhysRevE.66.066108} {\bibfield
  {journal} {\bibinfo  {journal} {Phys. Rev. E}\ }\textbf {\bibinfo {volume}
  {66}},\ \bibinfo {pages} {066108} (\bibinfo {year} {2002})}\BibitemShut
  {NoStop}%
\bibitem [{\citenamefont {Anzaki}\ \emph {et~al.}(2015)\citenamefont {Anzaki},
  \citenamefont {Fukushima}, \citenamefont {Hidaka},\ and\ \citenamefont
  {Oka}}]{anzaki2015restricted}%
  \BibitemOpen
  \bibfield  {author} {\bibinfo {author} {\bibfnamefont {R.}~\bibnamefont
  {Anzaki}}, \bibinfo {author} {\bibfnamefont {K.}~\bibnamefont {Fukushima}},
  \bibinfo {author} {\bibfnamefont {Y.}~\bibnamefont {Hidaka}},\ and\ \bibinfo
  {author} {\bibfnamefont {T.}~\bibnamefont {Oka}},\ }\href
  {https://doi.org/https://doi.org/10.1016/j.aop.2014.11.004} {\bibfield
  {journal} {\bibinfo  {journal} {Ann. Phys.}\ }\textbf {\bibinfo {volume}
  {353}},\ \bibinfo {pages} {107 } (\bibinfo {year} {2015})}\BibitemShut
  {NoStop}%
\bibitem [{\citenamefont {Ito}\ \emph {et~al.}(2019)\citenamefont {Ito},
  \citenamefont {Nagao}, \citenamefont {Kurokawa}, \citenamefont {Kasuya},\
  and\ \citenamefont {Inoue}}]{ito2019bayesian}%
  \BibitemOpen
  \bibfield  {author} {\bibinfo {author} {\bibfnamefont {S.}~\bibnamefont
  {Ito}}, \bibinfo {author} {\bibfnamefont {H.}~\bibnamefont {Nagao}}, \bibinfo
  {author} {\bibfnamefont {T.}~\bibnamefont {Kurokawa}}, \bibinfo {author}
  {\bibfnamefont {T.}~\bibnamefont {Kasuya}},\ and\ \bibinfo {author}
  {\bibfnamefont {J.}~\bibnamefont {Inoue}},\ }\href
  {https://doi.org/10.1103/PhysRevMaterials.3.053404} {\bibfield  {journal}
  {\bibinfo  {journal} {Phys. Rev. Mater.}\ }\textbf {\bibinfo {volume} {3}},\
  \bibinfo {pages} {053404} (\bibinfo {year} {2019})}\BibitemShut {NoStop}%
\end{thebibliography}%
\bibliographystyle{apsrev4-2}

\end{document}